\pdfoutput=1

\documentclass[pss]{wiley2sp}
\usepackage{amssymb}

\begin{document}

\titlefigure[width=0.25\textwidth]{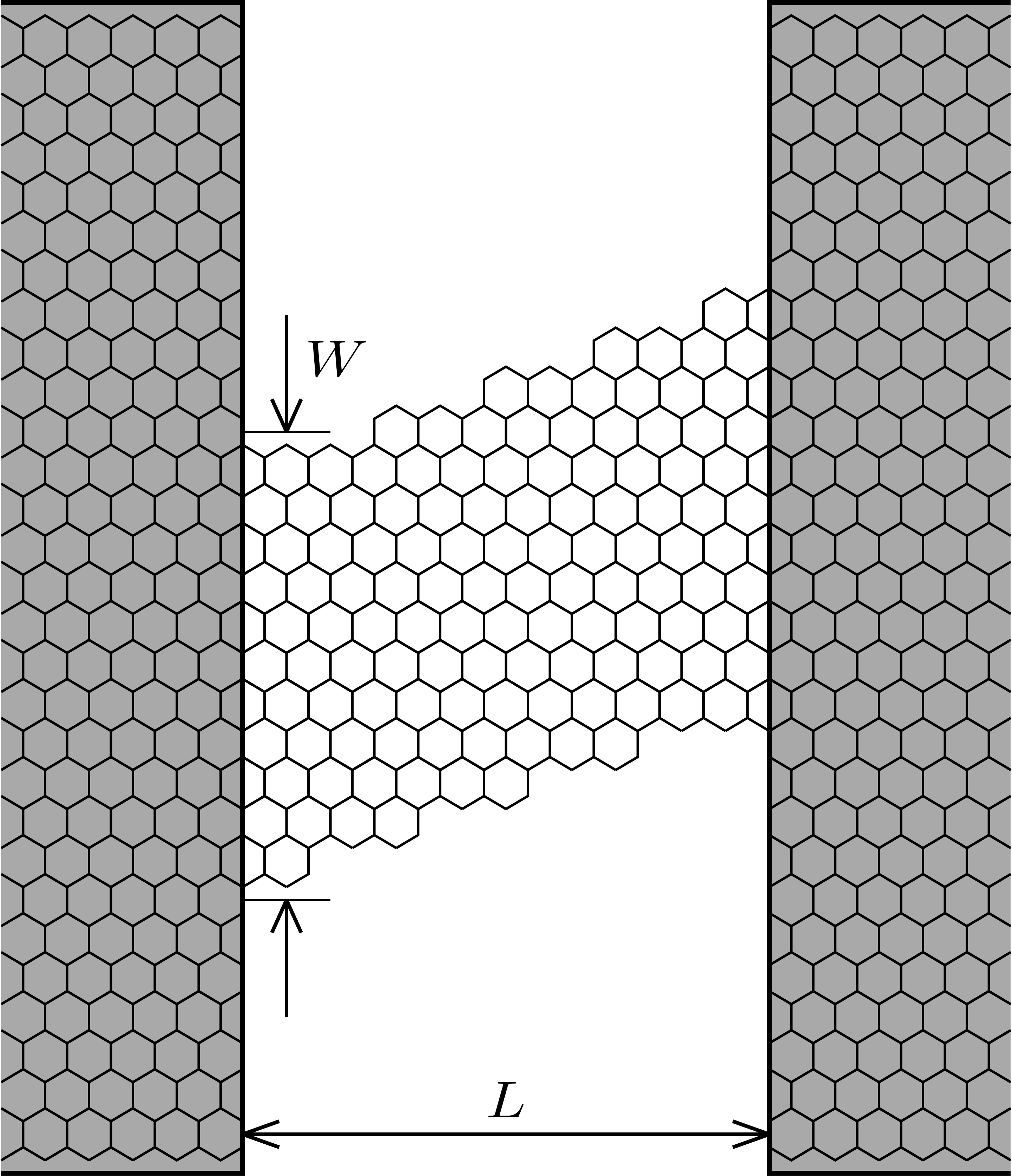}
\titlefigurecaption{%
\textbf{Generic valley filter} attached to heavily-doped rectangular graphene leads (shadow areas). Notice the constriction edges are neither perfect \emph{zigzag} nor \emph{armchair}.
}

\title{Nonequilibrium valley polarization in graphene nanoconstrictions}
\titlerunning{Nonequilibrium valley polarization in graphene nanoconstrictions}

\author{%
  Adam Rycerz\textsuperscript{\textsf{\bfseries 1,\Ast}}}

\authorrunning{A.\ Rycerz}

\mail{e-mail \textsf{rycerz@th.if.uj.edu.pl}, 
  Phone +48 12 663 5568, Fax +48 +12 633 4079}

\institute{%
  \textsuperscript{1}\,Marian Smoluchowski Institute of Physics, Jagiellonian University, Reymonta 4, 30-059 Krak\'{o}w, Poland}

\received{XXXX} 
\published{XXXX}


\abstract{%
\abstcol{%
We recently shown, using tight-binding calculations, that nonequilibrium valley polarization can be realized in graphene, when the current is injected through "valley filter": a ballistic point contact with zigzag edges. Here we demonstrate, that the effect is surprisingly robust against changing the crystallographic orientation of the filter axis. Namely, the output current remains polarized unless a point contact has perfect armchair edges, at which two subblattices are equally represented. The polarization is inverted when the filter orientation crosses the amchair line and, subsequently, dominating subblattice index of terminal atoms changes. In a bended graphene strip, the valley-polarized current can be directed towards one edge providing a theoretical possibility to observe a zero-magnetic-field analogue of the well-known Hall effect. For the valley valve, build of two valley filters in series and controlled elecrostatically by a gate voltage, the conductance-to-gate characteristic is inverted when $\pi/3$ vertex is placed between two filters. 
}{%
}}

\maketitle

\section{Introduction} 
Graphene, the new and probably the most intriguing one of carbon allotropes, have become a subject of intense research focus in last few years. As its low-energy spectrum is described by the Dirac-Weyl Hamiltonian of massless spin--$1/2$ fermions \cite{Geimreview}, graphene offers the unique possibility to test predictions of relativistic quantum mechanics in a condensed phase. However, in graphene, Dirac electrons have an additional degeneracy, corresponding to the presence of two different valleys in the band structure. This phenomenon, known as "fermion doubling", makes difficult to probe the \emph{intrinsic} physics of a single valley in experiments, because usually the contributions of the two valleys to a measurable quantity exactly cancels each other. An interesting example is production of a pseudo-mag\-ne\-tic field, which affects the electron in a single valley, by a lattice defect or distortion \cite{AMFG}. The field has an opposite sign in the other valley, so the effect remains a playground for theoreticians as long as both valleys are equally populated in an experiment. These are the fundamental reasons to look for a feasible and controlled way of the valley-isospin manipulation  in graphene. 

The practical consequences of eliminating fermion doubling effects are even more exciting: The spin-based quantum computing in graphene quantum dots \cite{graphenespinqubits}, exploiting the superior spin coherence expected in carbon nanostructures, is one of the most recent examples. More generally, the potential of graphene for future electronics rests on the possibility to create devices that have no analogue in silicon-based electronics \cite{Nov04}. Briefly speaking, it seems that valley degree of freedom might be used to control an electronic device, in much the same way as the electron spin is used in spintronics \cite{Wol01} or quantum computing \cite{Cer05}.

Therefore, a controllable way of occupying a single valley in graphene (in other words: producing a valley polarization) would be a key ingredient for the so-called ``valleytronics''. The author, Tworzyd{\l}o, and Beenakker recently proposed an electrostatic method of valley polarization in a single-mode quantum point contact with zigzag edges \cite{Ryc07a}. The work was build on earlier findings \cite{Fuj96,Nak96,Wak01,Wak02,Per06,Bre06,Two06} for graphene ribbons (long and narrow ballistic strips), that they may support a propagating mode arbitrarily close to the Dirac point, and that this mode lacks the valley degeneracy of modes that propagate at higher energies. For armchair edges of the ribbon, this lowest propagating mode is constructed from states in both valleys, but for zigzag edges only a single valley contributes \cite{Fuj96,Nak96,Wak01,Wak02,Per06}. In accord with time-reversal symmetry, the mode switches from one valley to the other upon changing the direction of propagation. We found \cite{Ryc07a}, that depending on whether the Fermi level in the point contact lies in the conduction or valence band, the transmitted electrons occupy states in one or the other valley of the band structure. Moreover, the effect is not limited to long and narrow ribbons, in fact the filter of the same length as width produces a polarization better than $95\%$. It was also shown in Ref.\ \cite{Ryc07a} that two adjacent valley filters function as a highly effective valley valve, passing or blocking the current depending on whether the two filters have the same or opposite polarity. Valley-filtering point contacts may also affect magnetoconductance oscillations in a multimode graphene ring, showing a suppression of the lowest harmonic for opposite polarity \cite{Ryc07b}.

Here we complement our previous study \cite{Ryc07a} by numerical simulation of the electron transport through a constriction with \emph{generic} edges, evolving from zigzag to armchair upon ``rotation'' of crystallographic orientation of the system axis. The results show that the output current lies predominantly in a single valley (chosen by a \emph{dominating} fraction of zigzag edges), unless the edges are perfectly of armchair type. A basic illustration of the valley-polarized current behavior in a bended-graphene strip is also provided. Finally, we show the function of valley filter (and valve) may be inverted by an appropriate edges reorientation. 

\begin{figure*}[!t]
\includegraphics*[width=\textwidth]{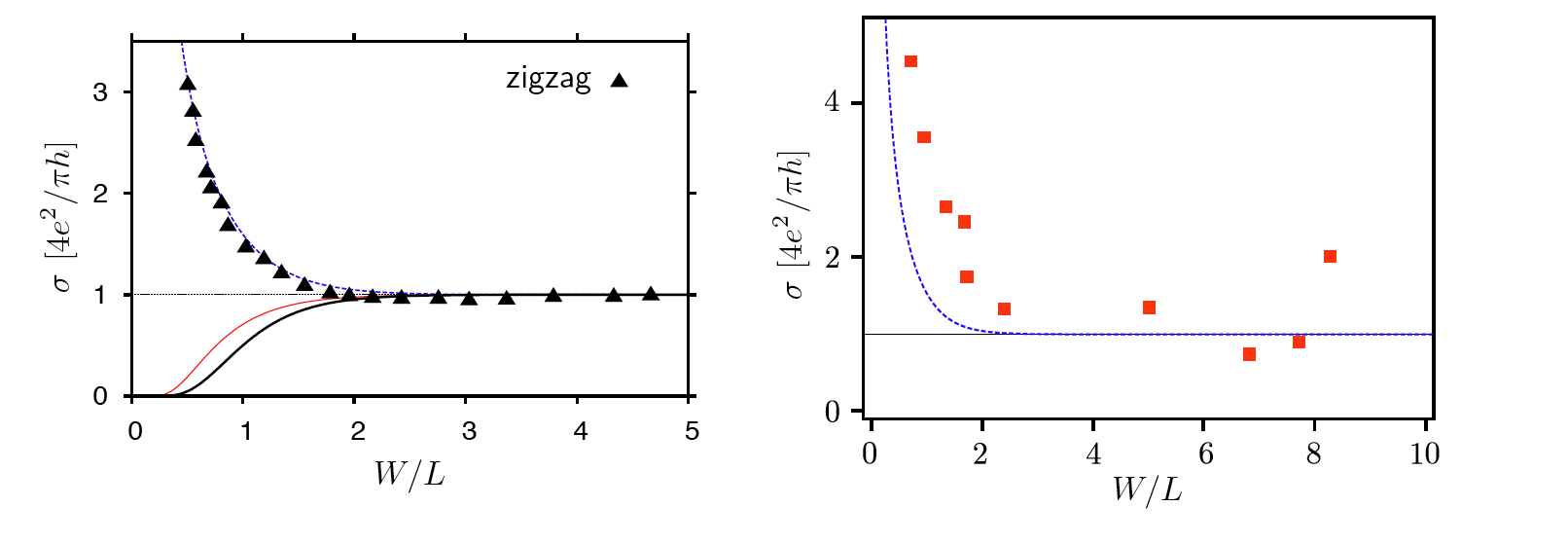}
\caption{\label{thexp}
Theory and experiment for a ballistic graphene monolayer. Left panel: conductivity at the Dirac point as a function of the aspect ratio of a graphene strip. The curves are calculated by solving the Dirac equation for different boundary conditions: infinite mass boudary condition (black tick line), semiconducting armchair edge (red thin line), and metallic armchair edge (blue dashed line). The datapoints for the zigzag edge (triangles) are the result of a numerical solution of the tight-binding model \cite{Tra06}. Right panel: Experimental results of Miao \emph{et al.}\ (red solid squares) compared with theory \cite{Miao:err} for metallic armchair edge (blue dashed line). 
}
\end{figure*}

\section{Preliminary: Ballistic transport in graphene}
In this section, we briefly recall the basic theoretical and experimental results concerning phase-coherent transport in gra\-phe\-ne at the Dirac point. First, we review the theory of Ref.\ \cite{Two06} by Tworzyd{\l}o \emph{et al.}, inspired by an insightful paper of Katsnelson \cite{Kat06}, who used the Landauer transmission formula to obtain the quantum-limited conductivity. Following the same approach, Tworzyd{\l}o \emph{et al.}\ calculated the transmission probabilities of Dirac particles through a rectangular strip of $W$ width and $L$ length, attached to heavily-doped graphene leads.

The low-energy excitations in graphene are govern by the massless Dirac equation
for the four-component wave function $\Psi=(\Psi_A,\Psi_B,\Psi'_A,\Psi'_B)$ in the general case (with $\Psi_X$, $\Psi'_X$ the amplitudes for different valleys, and $X=A,B$ the sublattice index). In the presence of external electrostatic potential $V(\mathbf{r})$ the wave equation may be written as
\begin{equation}
\label{dirac}
\frac{\hbar v}{i}\left(\begin{array}{cc}
 \sigma_x\partial_x+\sigma_y\partial_y & 0 \\
 0 & \sigma_x\partial_x-\sigma_y\partial_y \\
\end{array}\right)\Psi
  + eV\Psi=\epsilon\Psi,
\end{equation}
where the velocity $v\equiv\frac{1}{2}\sqrt{3}\tau a/\hbar\approx 10^6\mbox{ m/s}$ (with $a=0.246\,\mbox{nm}$ the lattice constant and $\tau=3.0\,\mbox{eV}$ the nearest-neighbor hopping) plays a role of the speed of light, $\sigma_i$ are the Pauli $2\times 2$ matrices, $e$ is the electron charge, and $\epsilon$ is the energy. We suppose the electrostatic potential $V=0$ in the strip ($0<x<L$) and $V=V_\infty$ elsewhere. In particular, the scattering problem for Eq.\ (\ref{dirac}) can be solved analytically for two different classes of boundary conditions \cite{AKBE}:
\begin{enumerate}

\item Infinite mass boundary condition
\begin{equation}
\left.\Psi\right|_{y=0}=
\left(\begin{array}{cc}
\sigma_x & 0 \\
0 & -\sigma_x \\
\end{array}\right)
\left.\Psi\right|_{y=0},
\end{equation}
\begin{equation}
\left.\Psi\right|_{y=W}=
\left(\begin{array}{cc}
-\sigma_x & 0 \\
0 & \sigma_x \\
\end{array}\right)
\left.\Psi\right|_{y=W}.
\end{equation}

\item Armchair edge
\begin{equation}
\left.\Psi\right|_{y=0}=
\left(\begin{array}{cc}
0 & 1 \\
1 & 0 \\
\end{array}\right)
\left.\Psi\right|_{y=0},
\end{equation}
\begin{equation}
\left.\Psi\right|_{y=W}=
\left(\begin{array}{cc}
0 & e^{i\phi} \\
e^{-i\phi} & 0 \\
\end{array}\right)
\left.\Psi\right|_{y=W},
\end{equation}
with $\phi=0$ for the \emph{metallic} armchair, and $\phi=\pm 2\pi/3$ the \emph{semiconducting} armchair (depending whether the strip width $W$ is, or is not an integer multiple of three unit cells).

\end{enumerate}

The results for conductance in a heavily-doped lead limit $|V_\infty|\rightarrow\infty$, including spin and valley degeneracies, may be summarized by \cite{Two06:app}
\begin{equation}
\label{stripg}
G=\frac{2e^2}{h}\sum_{n=-\infty}^{\infty}T_n,\ \ \ 
T_n=\frac{1}{\cosh^2[\pi(n+\alpha)L/W]},
\end{equation}
with $\alpha=1/2$ for infinite mass boundary conditions, $\alpha=0$ for the metallic armchair edge, and $\alpha=1/3$ for the semiconducting armchair edge. In the limit of wide and short strip $W\gg L$ the formula (\ref{stripg}) leads to the universal value of the conductivity
\begin{equation}
\label{stripsig}
\sigma=G\frac{L}{W}\approx \frac{4e^2}{\pi h},
\end{equation}
regardless the boundary conditions. The result (\ref{stripsig}) for the conductivity agrees with other theoretical calculations \cite{Per06,Lud94} which start from an unbounded disordered system and then take the limit of infinite mean free path $l$. For a long time, existing experiments \cite{Nov05,Zha05} were quasi-ballistic, with $l\sim W\sim L$, and finding $\sigma\approx 4e^2/h$. The ``missing $1/\pi$ factor'' was found very recently \cite{Mia07}, terminating the main controversy on ballistic graphene conductance. 

Ref.\ \cite{Mia07} provides also first experimental insight into the conductivity dependence on the aspect ratio $W/L$. In Fig.\ \ref{thexp} we compare the analytical results \cite{Two06}, reviewed in this section, with a computer simulation of electron transport for zigzag edge (where the analytical solution is lacking), and with the experimental results of Miao \emph{et al.} \cite{Mia07}. First, we found \cite{Tra06} that formula (\ref{stripg}) with $\alpha=0$ (metallic armchair) match precisely the numerical data for zigzag edges (see left panel of Fig.\ \ref{thexp}, blue dashed line and triangles, respectively). This is caused by the fact that there is a nondegenerated propagating mode at zero energy in both cases. (Notice also  the results for the semiconducting armchair are close to that for infinite mass boundaries, since there is no propagating mode there.) This finding strengthen the theory significantly, as the presence of metallic armchair edge requires the width to be an integer multiplicity of three unit cell and thus seems unrealistic in an experimental sample. Moreover, the recent findings \cite{Akhmerov:unpublished} show that the zigzag boundary condition applies generically for any angle $\theta\neq 0\;({\rm mod}\;\pi/3)$ of the boundary (where $\theta=0$ labels the armchair orientation). 

For the purposes of this paper, conductivity dependence on aspect ratio in the range $W/L<2$ is particularly important, as it indicates the degeneracy of the lowest propagating mode. The experimental results of Miao \emph{et al.}\ still lies above the theory \cite{Miao:err} in this range, with a relatively large disperion of datapoints (see right panel of Fig.\ \ref{thexp}). The probable explanation is related to an uncertainity in $W$ and $L$ for small samples, which may cause a systematic horizontal shift of the data. (Notice the conductivity increase rapidly when lowering $W/L$ below the threshold value $W/L\approx 2$.) One may also attribute the deviation between theory and experiment to the smooth (long-range) disorder, which was shown \cite{Nom07} to be responsible for enhanced condictivity in previous experiments. Albeit succesfully eliminated in work \cite{Mia07} for short and wide samples, the smooth disorder may still play some role for long and narrow samples, because the charge screening become less effective in such quasi-one-dimensional systems. Nevertheless, the remarkable experimental progress achieved by Miao \emph{et al.}\ allows us to believe, that ballistic graphene samples with $W/L<2$ and electric transport dominated by a single, valley-polarized mode, are to be fabricated shortly. The aim of this paper is computer modelling of the measurements directly demonstrating the valley-isospin manipulation in such nanostructures.

\begin{figure}[!p]
\includegraphics*[width=0.9\linewidth]{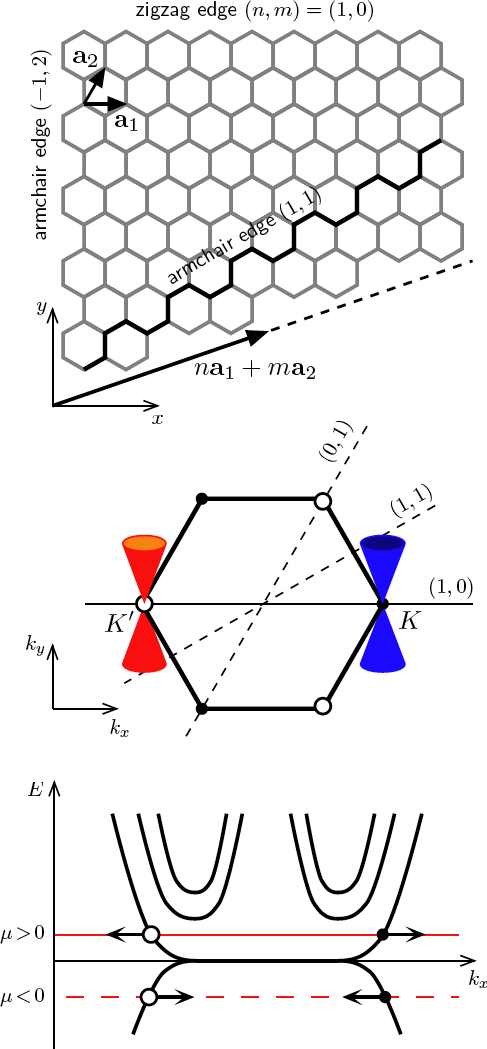}
\caption{\label{latti}
Schematic illustration of the mechanism of valley filtering in graphene nanoconstrictions. Top panel: Honeycomb lattice of a carbon monolayer with $\mathbf{a}_1=(a,0)$ and $\mathbf{a}_2=(\frac{1}{2},\frac{\sqrt{3}}{2})a$ basis vectors ($a$ is the lattice spacing). Generic edge of a finite system is defined by a pair of integers $(n,m)$. Middle panel: Hexagonal first Brillouin zone of graphene. The conically shaped valleys touch six corners of the Brillouin zone, the three marked by black (white) dots are equivalent. Bottom panel: Dispersion relation for the constriction with $(1,0)$ zigzag edges. For the lowest mode, there is one-to-one correspondence between the direction of propagation (indicated by arrows) and the valley isospin. The polarity switches $(i)$ when changes the sign of chemical potential $\mu$, or $(ii)$ when rotates the constriction axis by $\pi/3$ (the $(0,1)$ zigzag case). The valleys are equally mixed for armchair edges.
}
\end{figure}

\section{Generic valley filter in graphene}\label{filsec}

\begin{figure}[!t]
\includegraphics*[width=0.9\linewidth]{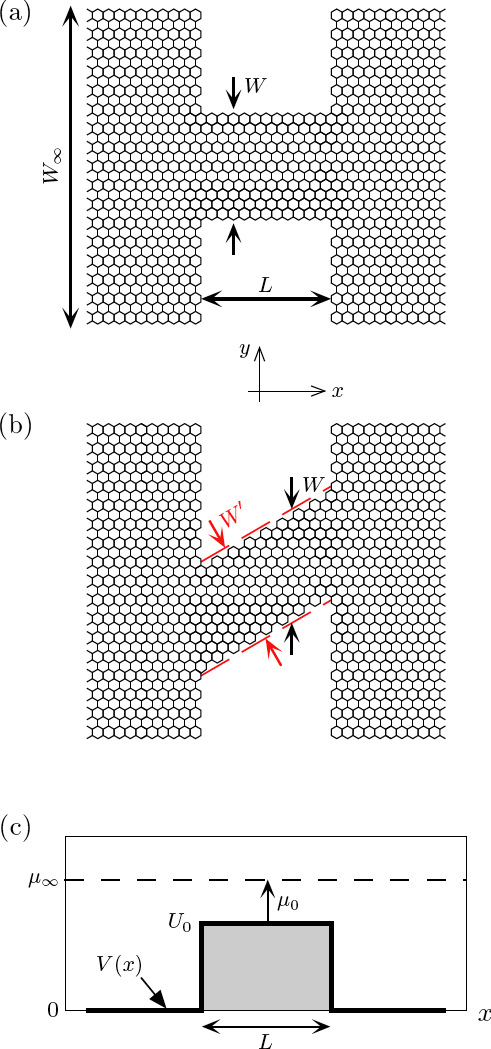}
\caption{\label{zachfi}
Schematic diagrams of valley filters with $(1,0)$ zigzag edges (a), $(1,1)$ armchair edges (b), and the corresponding potential profile $V(x)$ (c). Here, $W_\infty=17\sqrt{3}\,a$, $W=6\sqrt{3}\,a$, $L=12a$. (Width of the constriction with armchair edges is $W'=W\cos\,\pi/6=9a$.) Constrictions with \emph{predominantly} zigzag edges allows one to control the output valley polarization by varying the electrostatic potential $U_0$.
}
\end{figure}

Here we consider the honeycomb lattice of carbon atoms in a strip containing a short and narrow constriction with quantized conductance. We suppose the constriction edges are parallel, namely, the narrow region is terminated at the top and the bottom along the same crystallographic axis $n\mathbf{a}_1+m\mathbf{a}_2$ (with $\mathbf{a}_1$, $\mathbf{a}_2$ the basis vectors). Therefore, the edges are defined by specifying a pair of integers $(n,m)$ and the constriction width $W$. The different ways of terminating the honeycomb lattice are illustrated in Fig.\ \ref{latti}. (Fig.\ \ref{zachfi} shows two basic examples of constrictions with $(1,0)$ zigzag and $(1,1)$ armchair edges.) For a long constriction, where the transport become one-dimensional, the crystallographic orientation of edges $(n,m)$ determines the direction of propagation in the first Brillouin zone (see Fig.\ \ref{latti}). As the lowest propagating mode in such a nanoribbon $(i)$ lack the twofold degeneracy of higher modes, and $(ii)$ may be arbitrary close to the Dirac point \cite{Fuj96,Nak96,Wak01,Wak02,Per06,Bre06,Two06}, the crystallographic orientation $(n,m)$ also determines the valley polarization of current passing the constriction.

\begin{figure*}[!t]
\includegraphics*[width=\textwidth]{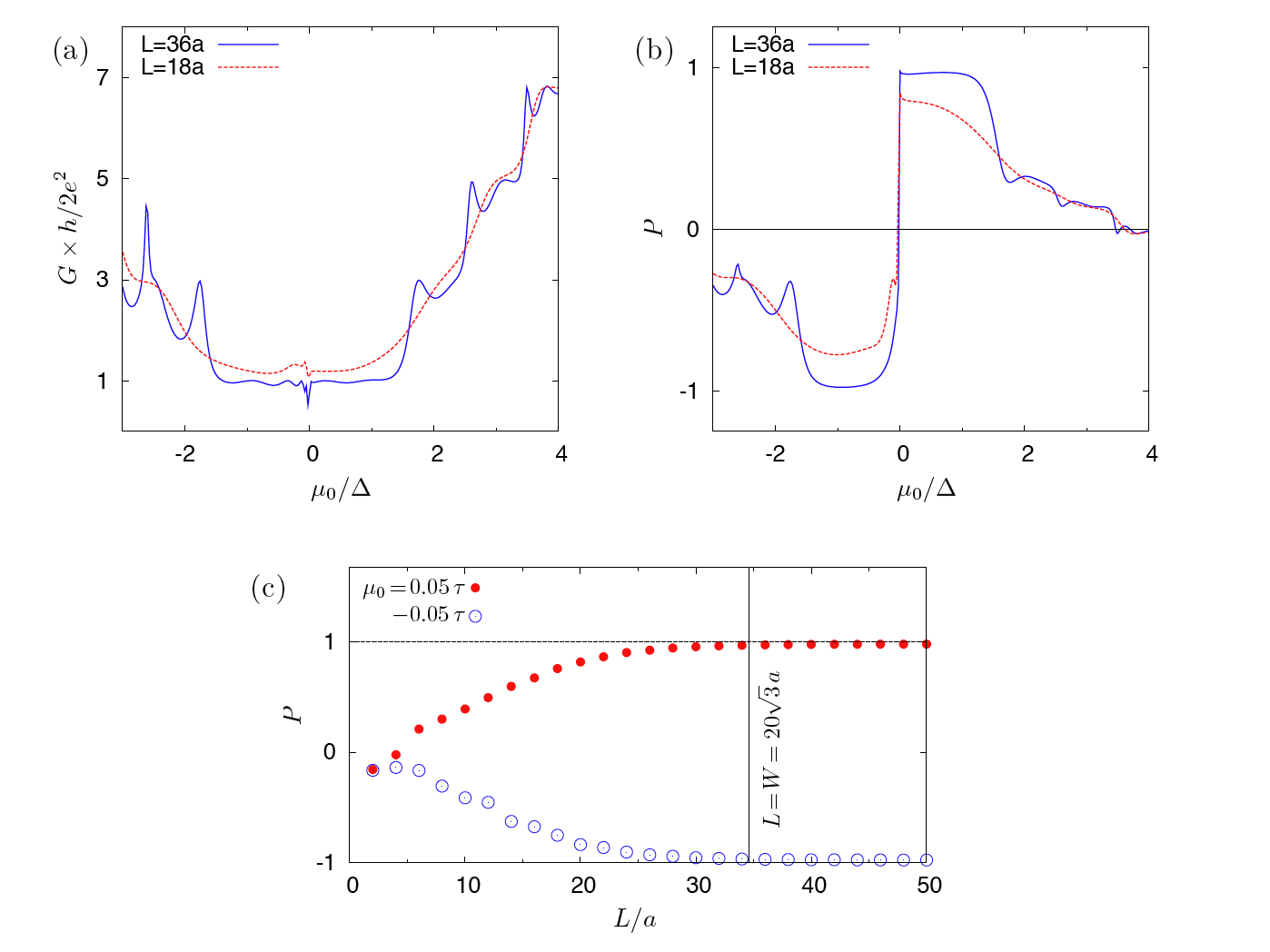}
\caption{\label{pfillen}
Operation of the valley filter with perfect \emph{zigzag} edges $(n,m)=(1,0)$, fixed width $W=20\sqrt{3}\,a$, and varying length $L$. Conductance (a) and valley polarization (b) for $L=36\,a$ (solid blue lines) and $L=16\,a$ (dashed red lines) as a function of $\mu_0/\Delta$ (with $\Delta=\pi\tau/40$). The parameter $\mu_0=\mu_\infty-U_0$ is varied by varying $U_0$ at fixed $\mu_\infty$. (c) The polarization at fixed $\mu_0=\pm 0.05\,\tau\approx\pm 2\Delta/3$ (solid red and open blue symbols, respectively) as a function the filter length $L$. Notice the saturation of $P$ for $L\gtrsim W$.}
\end{figure*}

\begin{figure*}[!t]
\includegraphics*[width=0.6\textwidth]{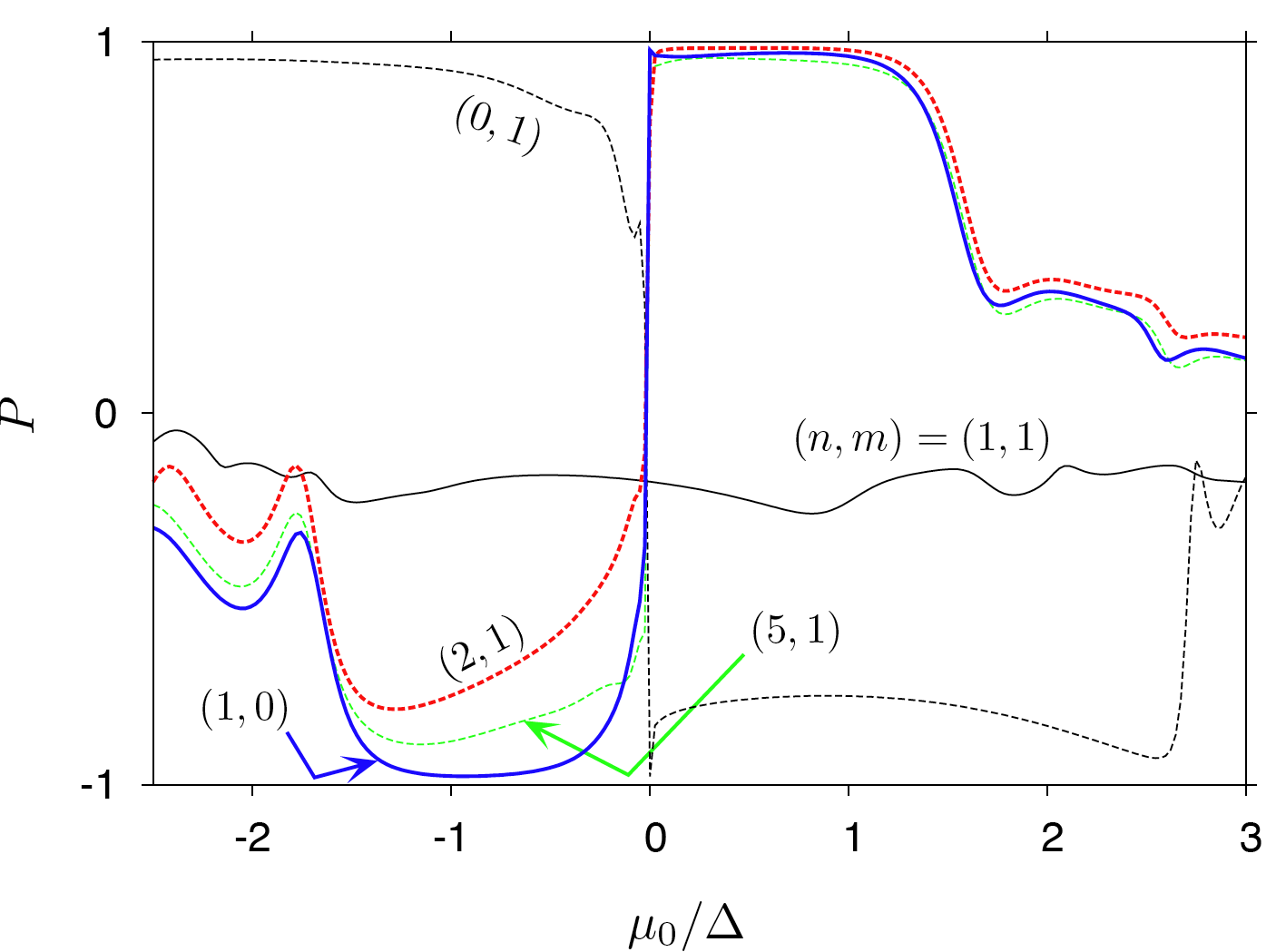}
\caption{\label{pfilind}
Valley polarization as a function of $\mu_0/\Delta$ and crystallographic orientation $(n,m)$ of the filter axis (specified for each curve on the plot).}
\end{figure*}

Our analysis starts from the tight-binding model of gra\-phe\-ne, with Hamiltonian
\begin{equation}
\label{hami}
  H=\sum_{i,j}\tau_{ij}|i\rangle\langle j|+\sum_{i}V_{i}|i\rangle\langle i|.
\end{equation}
The hopping matrix element $\tau_{ij}=-\tau$ if the orbitals $|i\rangle$ and $|j\rangle$ are nearest neighbors on the honeycomb lattice, otherwise $\tau_{ij}=0$. $V_{i}=V(x_{i})$ varies only along the main axis of a strip (see Fig.\ \ref{zachfi}). Namely, the potential equals $U_0$ inside the constriction ($x<0<L$, where $L$ is the constriction length) and zero everywhere else. By varying $U_0$ at a fixed Fermi energy $\mu_\infty$ in the external leads (wide regions), we can vary the Fermi energy $\mu_0=\mu_\infty-U_0$ inside the constriction. 

The wide regions in Fig.\ \ref{zachfi} have $(1,0)$ zigzag edges, and therefore support $2N+1$ propagating modes at the Fermi energy $\mu_\infty$, which form a basis for the transmission matrix $t$. Modes $l=1,2,\ldots N$ lie in the first valley [with longitudinal wave vector $k_xa\in(\pi,2\pi)$] while modes $l=-1,-2,\ldots -N$ lie in the second valley [with $k_xa\in(0,\pi)$]. The zeroth mode $l=0$ lies in a single valley fixed by the direction of propagation. The conductance of the constriction is determined by the Landauer formula
\begin{equation}
\label{Landauer}
G=\frac{2e^{2}}{h}\sum_{l=-N}^{N}T_{l},\ \ T_{l}=\sum_{k=-N}^{N}|t_{lk}|^{2}.
\end{equation}
The valley polarization of the transmitted current is quantified by
\begin{equation}
\label{Pdef}
P=\frac{T_{0}+\sum_{l=1}^{N}(T_{l}-T_{-l})}{\sum_{l=-N}^{N}T_{l}},
\end{equation}
where we consider the case $\mu_\infty>0$, such that the zeroth mode lies in the first valley. The polarization $P\in[-1,1]$, with $P=1$ if the transmitted current lies fully in the first valley and $P=-1$ if it lies fully in the second valley. Throughout the paper, we took $\mu_\infty=\tau/3$ to work with heavily doped graphene leads, while remaining at sufficiently small Fermi energy that the linearity of the dispersion relation holds reasonably well.

We have calculated the transmission matrix numerically by adapting to the honeycomb lattice the method developed by Ando for a square lattice \cite{And91}. Results are shown in Figs.\ \ref{pfillen} and \ref{pfilind}. We have fixed the width of the wide regions at $W_{\infty}=70\sqrt{3}\,a$, corresponding to $2N+1=29$ propagating modes. The narrow region has width $W=20\sqrt{3}\,a$ (measured along $y$-axis, see Fig.\ \ref{zachfi}a,b). We measure the electrochemical potential $\mu_{0}$ in the narrow region in units of the subband splitting for $(1,0)$ zigzag edge strip
\begin{equation}
\label{delta}
\Delta\equiv\frac{1}{2}\sqrt{3}\,\pi\tau a/W = \pi\hbar v/W.
\end{equation}
For our parameters $\Delta=\pi\tau/40$.

In Fig.\ \ref{pfillen} we demonstrate the operation of the valley filter with $(1,0)$ zigzag edges. Figs.\ \ref{pfillen}a and \ref{pfillen}b shows the conductance and valley polarization, both as a function of the electrochemical potential $\mu_{0}$ in the narrow region. For positive $\mu_{0}$ the current flows entirely within the conduction band, and we obtain plateaus of quantized conductance at odd multiples of $2e^{2}/h$ (as predicted by Peres et al.\ \cite{Per06}). The plateaus in the conductance at $G=(2n+1)\times 2e^{2}/h$ correspond to plateaus in the valley polarization at $P=1/(2n+1)$. For negative $\mu_{0}$ the current makes a transition from the conduction band in the wide regions to the valence band in the narrow region.  The interband transition destroys the conductance quantization --- except on the first plateau, which remains quite flat in the entire interval $-3\Delta/2<\mu_{0}<3\Delta/2$. The resonances at negative $\mu_{0}$ are due to quasi-bound states in the valence band \cite{Mil06,Sil06}. The polarity of the valley filter is inverted for negative $\mu_{0}$, almost without loosing the quality. 

The data for quadratic filter $L=36a\approx W$ (blue solid curves in Figs.\ \ref{pfillen}a and \ref{pfillen}b) attached to \emph{rectangular} leads (cf.\ Fig.\ \ref{zachfi}a) are almost identical with those obtained for the filter with \emph{triangular} leads, studied in Ref.\ \cite{Ryc07a}, showing the lead-effects are negligible as the transport through the device is determined by a constriction. The new insight into the nature of valley polarization is provided with the data for shorter filters. For $L=16a\approx W/2$ (red dashed curves) the polarization is still above $70\%$ for $|\mu_0|\lesssim\Delta$ (in comparizon to $95\%$ for $L\approx W$), although the conductance plateaus are smeared out. This is because the contribution to the conductance from evanescent and propagating modes  become coparable for $W/L\gtrsim 2$ (notice the conductivity in Fig.\ \ref{thexp} is not far from the universal limit in this range). In contrast, the contributions from evenescent modes in two valleys cancel each other in Eq.\ (\ref{Pdef}), expaining the robustness of this quantity when changing the aspect ratio. (The fast saturation of the polarization with increasing $L$ is illustrated in Fig.\ \ref{pfillen}c for two selected values of the chemical potential $\mu_0=\pm 0.05\,\tau\approx\pm 2\Delta/3$.) Therefore, it would be very benefitial to find a quantity, which is proportional to $P$ and, simultaneously, feasible to measure in an experiment. One of the candidates is the transverse Hall voltage induced by a pseudo-magnetic field in bended graphene strip \cite{AMFG}, briefly discussed in the next section.

\begin{figure*}[!t]
\includegraphics*[width=0.9\textwidth]{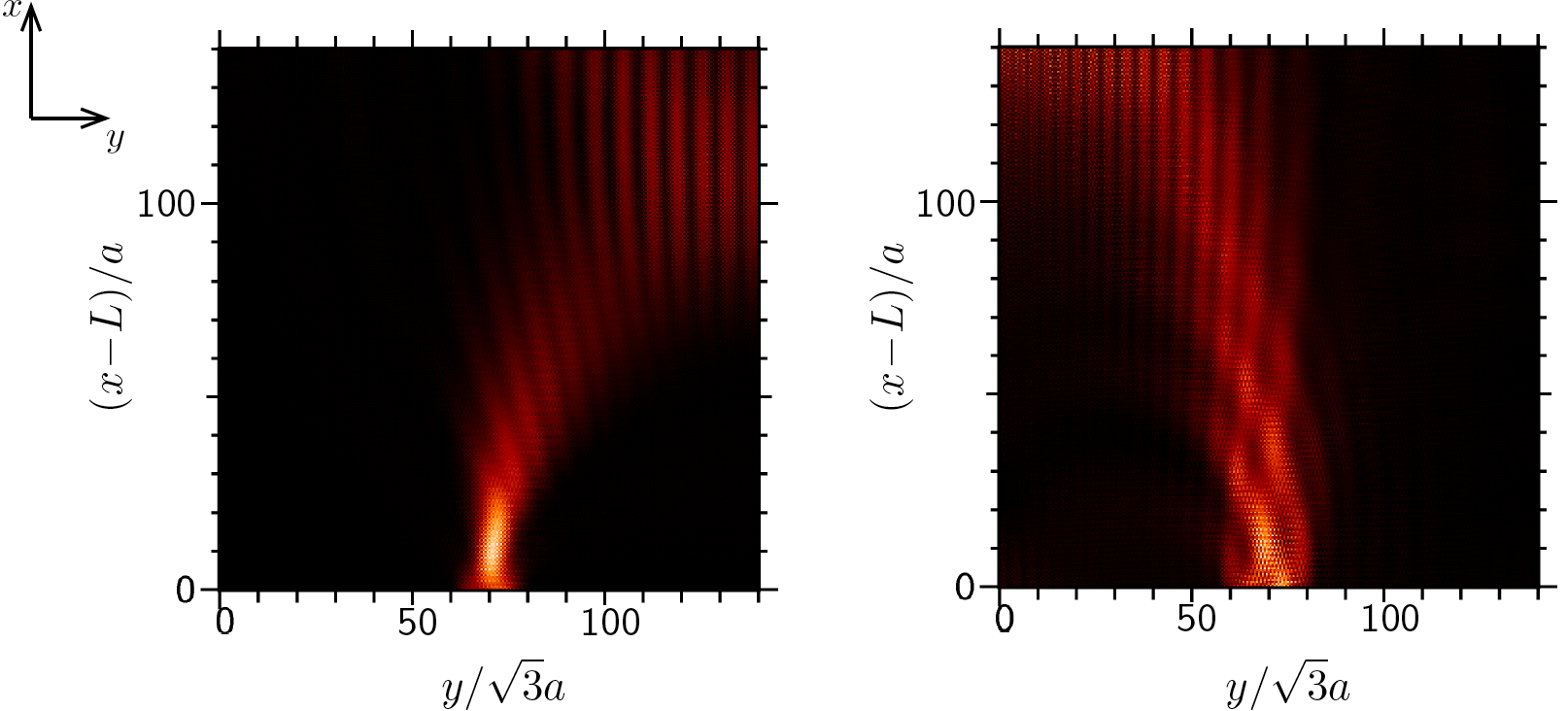}
\caption{\label{benbe}
Probability distribution $|\Psi_\mathrm{out}(\mathbf{x})|^2$ for the electron after passing the valley filter (not shown) in a bended graphene strip. Left and right panel show the cases of an opposite chemical potential inside the constriction $\mu_0=\pm 0.05\,\tau$. The remaining filter parameters are $W=20\sqrt{3}a$, $L=34a$, and $(n,m)=(1,0)$. The coordinate system of Fig.\ \ref{zachfi} is also shown.
}
\end{figure*}

But first, we report the principal new results presented in Fig.\ \ref{pfilind}, where we show the evolution of valley polarization when changing the filter crystallographic orientation (specifies by $(n,m)$ indices). For $m/n<1$ the polarization is almost unaffected for $\mu_0>0$, and changes gradually with increasing $m/n$ for $\mu_0<0$. Again, the polarization is more than $70\%$ in wide parameter range. This is because the large Fermi wavelength at small $\mu_0$ causes the quality of the valley filter is quite robust against edge imperfections, like a nonzero fraction of armchairs separting the perfect zigzags. For $n=m$ (the perfect armchair edges) we obtain $P\approx 0$, as the valleys are equally almost mixed in this case. For $(0,1)$ zigzag edge case the polarization is inverted, as it corresponds to the rotation of crystalographic axis by $\pi/3$. The filter behavior is therefore consistent with the recent theory of Ref.\ \cite{Akhmerov:unpublished}, which has shown that the zigzag boundary condition applies generically to any crystallographic orientation of the lattice except the \emph{exact} armchair orientation.

\section{Valley polarized current in bended graphene}\label{bensec}
To illustrate the effect of pseudo-magnetic field \cite{AMFG} in a bended graphene strip we modify the off-diagonal part in Hamiltonian (\ref{hami}). The vertical hopping elements varies now with $y$ position across the strip (see Fig.\ \ref{zachfi}), namely, for $|i\rangle$ and $|j\rangle$ the nearest neighbors
\begin{equation}
\tau_{ij}=\left\{ 
\begin{array}{cl}
-\tau(1+\Theta y_i), & \mbox{if}\ \ x_i=x_j \\
-\tau, & \mbox{if}\ \ x_i\neq x_j\\
\end{array}.
\right.
\end{equation}
This corresponds to the uniform perpendicular pseudo-mag\-ne\-tic field $\Theta$ (we use the same pseudo-field in both leads as well as inside the constriction). Here, we took $\Theta W_\infty=0.5$, with $W_\infty=140\sqrt{3}a$. 

 In Fig.\ \ref{benbe} we plot the probability distribution in the output lead $|\Psi_{\mathrm{out}}(\mathbf{x})|^2$ (with $\mathbf{x}=(x,y)$ the position on a honeycomb lattice \cite{nospino}), which is summarized over all incomming modes, each one contributing with
\begin{equation}
  |\Psi_\mathrm{out}^{(l)}(\mathbf{x})|^2 = 
  \left|\sum_k t_{kl}\phi_k(\mathbf{x})\right|^2=
\end{equation}
$$
  \sum_{kk'}t_{kl}^\star t_{k'l}\phi_k^\star(\mathbf{x})\phi_{k'}(\mathbf{x}),
$$
where $\phi_{k}(\mathbf{x})$ is the wave function of $l$-th mode (note we consider an \emph{identical} input and output leads). The compact formula for $|\Psi_{\mathrm{out}}(\mathbf{x})|^2$ reads
\begin{equation}\label{psiout2}
|\Psi_{\mathrm{out}}(\mathbf{x})|^2=\sum_l|\Psi_\mathrm{out}^{(l)}(\mathbf{x})|^2 = 
\mbox{Tr}\left(tt^\dagger\rho[\mathbf{x})\right],
\end{equation}
where $\rho(\mathbf{x})$ is the density operator with matrix elements $[\rho(\mathbf{x})]_{kk'}=\phi_k^\star(\mathbf{x})\phi_{k'}(\mathbf{x})$. 

The data presented in Fig.\ \ref{benbe} shows the current leaving valley filter forms a well colimated beam, particulary in the case of positive polarization (left panel), when the electron stays in a conductions band when passing the constriction. (The cyclotronic length in both cases is $l_\Theta=\mu_\infty/(\sqrt{3}\tau\Theta)\approx 100a$.) The effect of bending is clearly opposite for the two valleys, suggesting a nonzero Hall voltage, proportional to the valley polarization $P$ in the lowest order, may appear at zero field in a bended graphene strip containing the constriction.

\begin{figure*}[!p]
\includegraphics*[width=0.91\textwidth]{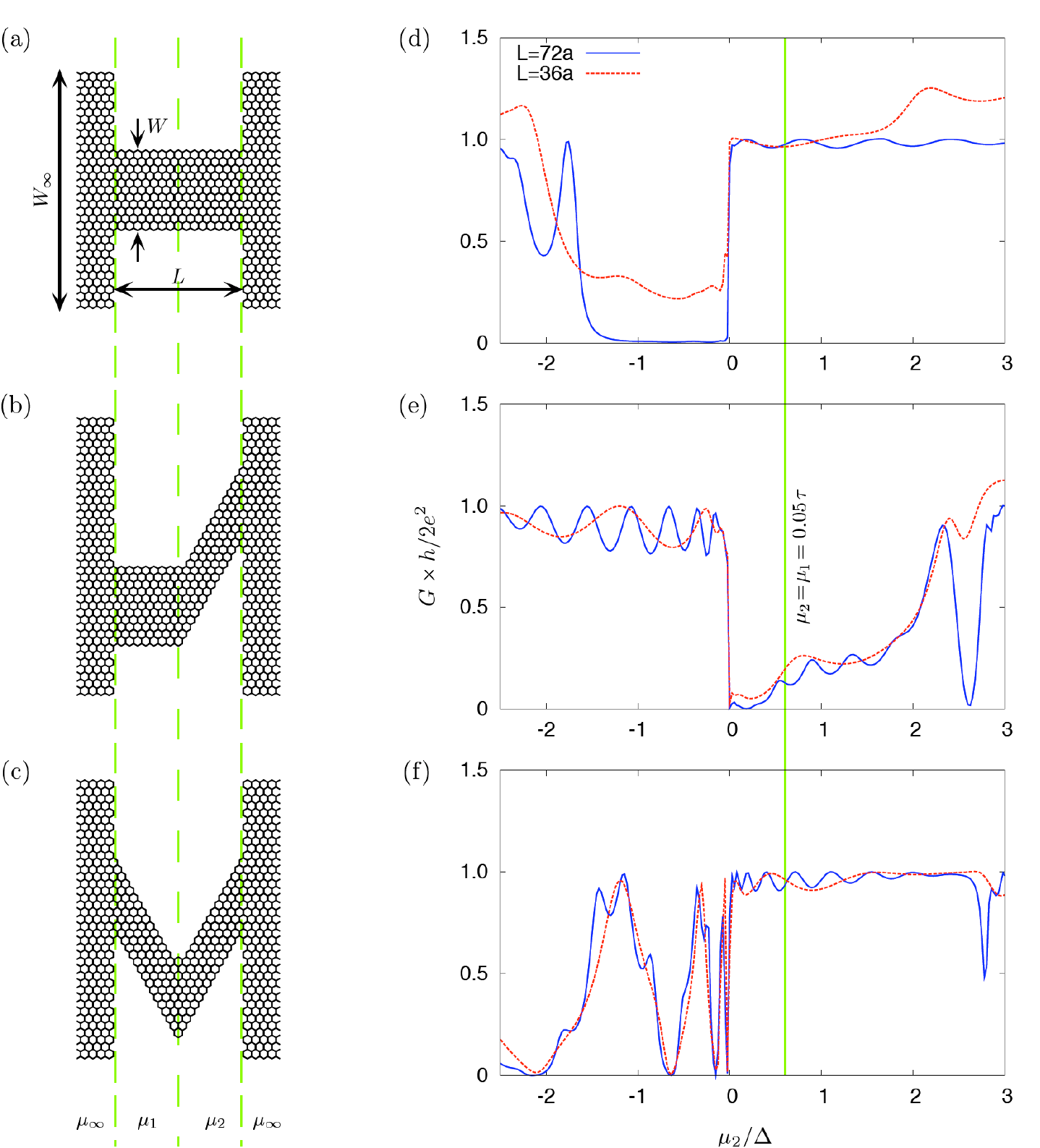}
\caption{\label{gvolver}
Different valley valves with zigzag edges (a--c) and their conductance (d--f) at fixed $\mu_1=0.05\,\tau\approx 2\Delta/3$ as a function of $\mu_2/\Delta$. Green dashed lines at diagrams (a--c) mark the electrochemical potential steps between values indicated at the bottom. Blue solid and red dashed curves at conductance plots (d--f) correspond to $L=72\,a$ and $36\,a$, respectively. Green vertical line show the case $\mu_1=\mu_2$ of no potential step between two valley filters building each valley valve.}
\end{figure*}

\section{Valley valve with a vertex}

Following the idea of Ref.\ \cite{Ryc07a}, we put now to different valley filters series, to show the conductance of the system is $G\approx 2e^2/h$ or $\approx 0$, depending whether the two filters have the same or opposite polarity. We analyze three systems, depicted schematically in Fig.\ \ref{gvolver}: first, build of two identical filters with zigzag edges (see Fig.\ \ref{gvolver}a), second with crystallographic axis of one filter rotated by $\pi/3$ (see Fig.\ \ref{gvolver}b), and third, consists of two rotated filters building a nanoribbon with $2\pi/3$ vertex in center (see Fig.\ \ref{gvolver}c). Thus, the current through each of these \emph{valley valves} have to change its direction by $\theta=0$, $\pi/3$, and $2\pi/3$ (respectively) when passing the center of a constriction. We use the lead width $W_\infty=140\sqrt{3}\,a$ (twice as large as in Section \ref{filsec}, to make sure the constriction is placed far away from lead edges) corresponding to $2N+1=59$ propagating modes in wide regions. The constriction width is $W=20\sqrt{3}\,a$ (defined along the transverse direction, as in Fig.\ \ref{zachfi}). Again, we perform the computations for two values of the constriction length $L/a=36$ and $72$. (So the valve is always build of two valley filters selected from those analyzed in Section \ref{filsec}.)

The results are shown in Figs.\ \ref{gvolver}d--f, where we plot the conductance at fixed chemical potential in the first filter $\mu_1=0.05\,\tau$ as a function of the chemical potential in the second filter $\mu_2$. The striking feature is an almost perfect inversion of the valve function each time when changing the relative angle $\theta$ between crystallographic axes of the two filters in the steps of $\pi/3$ (with some loose of the quality for $\mu_2<0$ in $\theta=2\pi/3$ case, where the complicated romboidal shape of a hole-doped quantum dot produce a relatively dense spectrum of resonant states). In particular, for the situation without a potential step between the filters $\mu_1=\mu_2=0.05\,\tau$ (indicated by green solid line along the right panel of Fig.\ \ref{gvolver}) the current is passed for $\theta=0$, $2\pi/3$ and blocked for $\pi/3$. In other words, the valve function is reduced in this case to the mechanism of current blocking at a vertex of a graphene hexagon with zigzag edges, discussed in Ref.\ \cite{Ryc07b}.

\section{Conclusions}
We have analyzed the operation on a valley filters with different crystallographic orientation on a honeycomb lattice. In a general case, the filter behavior is dominated by a zigzag line that is closest to its actual edge direction. The function of a filter is inverted when passing the armchair orientation. The effect of pseudomagnetic field on a valley-polarized current in bended graphene have been also demonstarted.

The conductance calculation for a valley valve shows it remains highly effective when changing the crystallographic orientation of the one of two filters in series. The valve function is insensitive on whether the same (or opposite) polarity of the filters is achieved electrostatically or by changing the geometry. 
These findings constitutes an alternative line of approach towards the practical realization of key ingredients for valleytronics.

\section*{Acknowledgment}\
This work was supported by a special grant of the Polish Science Foundation (FNP) and by the Polish Ministry of Science (Grant No. 1--P03B--001--29). Discussions with Patrik Recher, Izak Snyman, Bj\"{o}rn Trauzettel, and Jakub Tworzyd{\l}o are gratefully acknowledged. I thank Anton Akhmerov for communicating to me the results of his theory \cite{Akhmerov:unpublished} before publication. A special thanks are addressed to Prof.\ C.W.J.\ Bee\-nak\-ker for many discussions, correspondence, and critical remarks on the manuscript.


\begin{thebibliography}{[1]}
\bibitem{Miao:err} Ref.\ \cite{Mia07} reports a much better agreement between theory and experiment than shown in Fig.\ 1, but it uses a formula for the conductance $G=(4e^{2}/h)\sum_{n=0}^{\infty}\cosh^{-2}(\pi nL/W)$ which incorrectly assumes twofold valley degeneracy for the lowest, fully transmitted, mode. Instead, this mode is nondegenerate for either zigzag or metallic armchair boundary conditions \cite{Two06,Two06:app}. In Fig.\ \ref{thexp} we compare the experiment with the conductance formula $G=(2e^{2}/h)\sum_{n=-\infty}^{\infty}\cosh^{-2}(\pi nL/W)$ of Eq.\ (\ref{stripg}) which avoids double counting of the lowest mode.
\bibitem{nospino}
We are not using the spinor notation of Eq.\ (\ref{dirac}) to underline the formula (\ref{psiout2}) for the probability distribution $|\Psi_\mathrm{out}(\mathbf{x})|^2$ applies to a scattering problem for any tight-binding model.
\end{thebibliography}
\end{document}